\documentclass[10pt,twocolumn,letterpaper]{article}

\usepackage{iccv}
\usepackage{times}
\usepackage{epsfig}
\usepackage{graphicx}
\usepackage{amsmath}
\usepackage{amssymb}
\usepackage{multirow}
\usepackage{multicol}

\usepackage{booktabs}
\usepackage[accsupp]{axessibility}  

\usepackage[pagebackref=true,breaklinks=true,letterpaper=true,colorlinks,bookmarks=false]{hyperref}

\graphicspath{{figures/}}

\usepackage[breaklinks=true,bookmarks=false]{hyperref}

\usepackage[capitalize]{cleveref}
\crefname{section}{Sec.}{Secs.}
\Crefname{section}{Section}{Sections}
\Crefname{table}{Table}{Tables}
\crefname{table}{Tab.}{Tabs.}

\iccvfinalcopy 

\def\iccvPaperID{1742} 

\ificcvfinal\fi

\begin{document}

\title{Unsupervised Image Denoising in Real-World Scenarios via Self-Collaboration Parallel Generative Adversarial Branches}

\author{Xin Lin\\
Sichuan University\\
{\tt\small linxin@stu.scu.edu.cn}
\and
Chao Ren\thanks{Corresponding Author}\\
Sichuan University\\
{\tt\small chaoren@scu.edu.cn}
\and
Xiao Liu\\
Sichuan University\\
{\tt\small liux@stu.scu.edu.cn}
\and
Jie Huang\\
Sichuan University\\
{\tt\small huangjiechn@stu.scu.edu.cn}
\and
Yinjie Lei\\
Sichuan University\\
{\tt\small yinjie@scu.edu.cn}
}

\maketitle
\ificcvfinal\thispagestyle{empty}\fi

\begin{abstract}
Deep learning methods have shown remarkable performance in image denoising, particularly when trained on large-scale paired datasets. However, acquiring such paired datasets for real-world scenarios poses a significant challenge. Although unsupervised approaches based on generative adversarial networks (GANs) offer a promising solution for denoising without paired datasets, they are difficult in surpassing the performance limitations of conventional GAN-based unsupervised frameworks without significantly modifying existing structures or increasing the computational complexity of denoisers. To address this problem, we propose a self-collaboration (SC) strategy for multiple denoisers. This strategy can achieve significant performance improvement without increasing the inference complexity of the GAN-based denoising framework. Its basic idea is to iteratively replace the previous less powerful denoiser in the filter-guided noise extraction module with the current powerful denoiser. This process generates better synthetic clean-noisy image pairs, leading to a more powerful denoiser for the next iteration. In addition, we propose a baseline method that includes parallel generative adversarial branches with complementary ``self-synthesis" and ``unpaired-synthesis" constraints. This baseline ensures the stability and effectiveness of the training network. The experimental results demonstrate the superiority of our method over state-of-the-art unsupervised methods. \url{https://github.com/linxin0/SCPGabNet}
\end{abstract}

\section{Introduction}
\label{sec:intro}
Image denoising aims to recover noise-free images from noisy observations by reducing the potential noise. Although it is one of the oldest and most classical tasks in low-level computer vision, its fundamental nature continues drawing much interest. In general, existing image denoising algorithms can be divided into three groups:
filtering-based \cite{2, 3, 4, 5, 66}, model-based methods \cite{ 9, 32, 30, 12, 69, 61} and learning-based \cite{17, 19, 21, 41, 36, 40, 44, 39, 38, 25} methods.
\begin{figure}[t]
\centering
\includegraphics[width=1\linewidth]{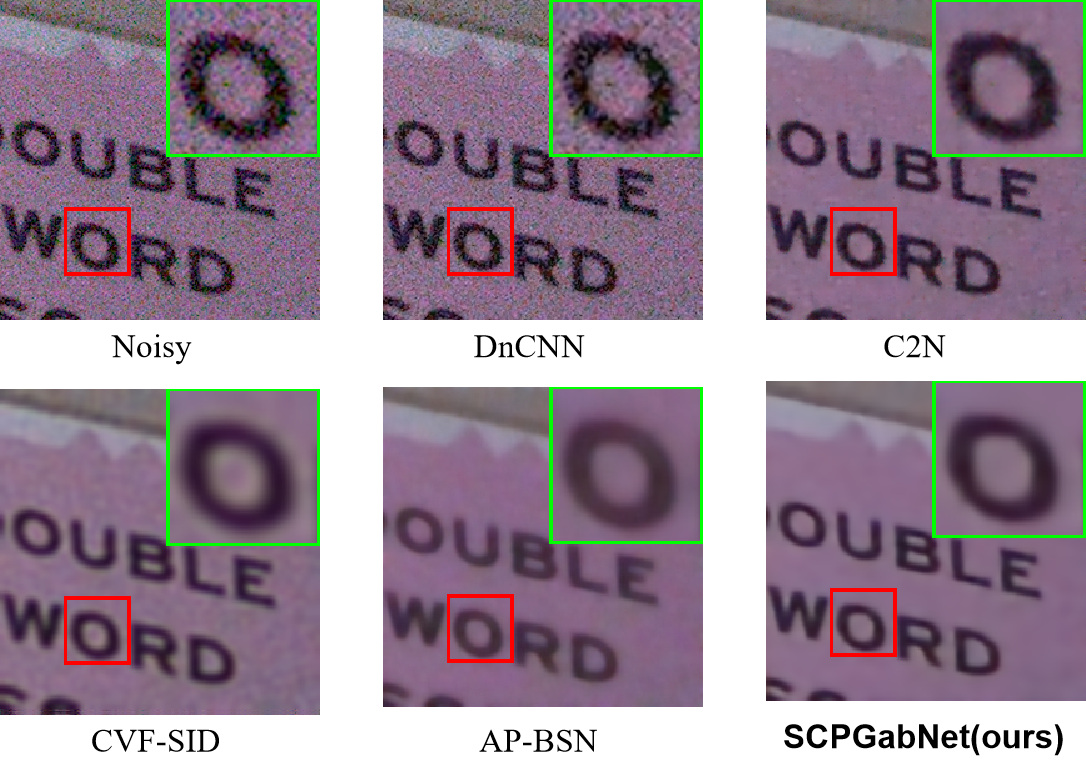}
\caption{\label{Fig3} A real noisy image from the SSID Validation dataset. Our SCPGabNet achieves better results compared to other denoising methods.}
\end{figure}
\par The additive white Gaussian noise (AWGN) assumption is widely used in image denoising. However, it is complex and challenging to adaptively achieve denoising based on filtering-based or model-based methods for high performance. In contrast, learning-based methods have demonstrated their superiority in image denoising. However, these methods \cite{54, 21, 25, 53, aind1, 100} are data-driven and typically require pairs of clean-noisy datasets to train their models. These noise samples are usually obtained through a predefined AWGN formulation that assumes the noise is signal-independent.
\par On the contrary, the real-world noise is more complex and much different from the ideal AWGN assumption. Using the AWGN model directly for the real scenes leads to poor performance. Therefore, numerous methods \cite{37, 40, 38, 39, 74, 90, aind1} have been proposed to capture paired clean-noisy image datasets from real scenes to promote the training of deep networks. However, these paired image-based methods focus on enhancing performance by improving network structures, and acquiring well-aligned pairs of clean-noisy images is time-consuming and laborious.
\par To solve the aforementioned problems, unsupervised denoising-based methods \cite{chen, hong, 1000, dbsnl} have emerged. The existing approaches are typically based on generative adversarial network (GAN) frameworks, which mainly focus on generating higher quality pseudo-noisy images. GAN2GAN \cite{cha} identified one of the key limitions of unsupervised denoising frameworks is the gap between the real and synthetic images, and proposed a novel approach using multiple generators and discriminators to generate images that closely conform to the real noise distribution. However, the performance of existing unsupervised denoising frameworks remains unsatisfactory due to the difficulty of adversarially training. Moreover, after training the model, the existing frameworks cannot further maximize the denoising potential without significantly changing its structure or increasing the inference complexity (e.g., using certain self-ensemble strategy) for denoisers. To address the previous limitations, we innovatively propose an unsupervised real-world denoising network called Self Collaboration Parallel Generative Adversarial Branches (SCPGabNet). The self-collaboration (SC) strategy, which provides the framework a powerful self-boosting capability. This enables the denoisers obtained from the conventional GAN framework to continuously evolve themselves and significantly improve their performance. The major contributions of our method are as follows:
\begin{itemize}
	\item[$\bullet$] We design a novel filter-guided synthetic noisy image generator with the noise extraction (NE) module to synthesize high-quality clean-noisy image pairs, which serve as the foundation for implementing the SC strategy.
	\item[$\bullet$] We propose an effective parallel generative adversarial branches framework with complementary ``self-synthesis" and ``unpaired-synthesis" constraints as our powerful baseline.
  \item[$\bullet$] We are the first to propose the SC strategy, which significantly enhances the performance of the GAN-based denoising framework without increasing its inference complexity. Experimental results demonstrate the superiority of our SCPGabNet over state-of-the-art unsupervised image denoising methods with large margins on the SIDD and DND benchmarking datasets.
\end{itemize}
\begin{figure*}[t]
\includegraphics[width=1\linewidth]{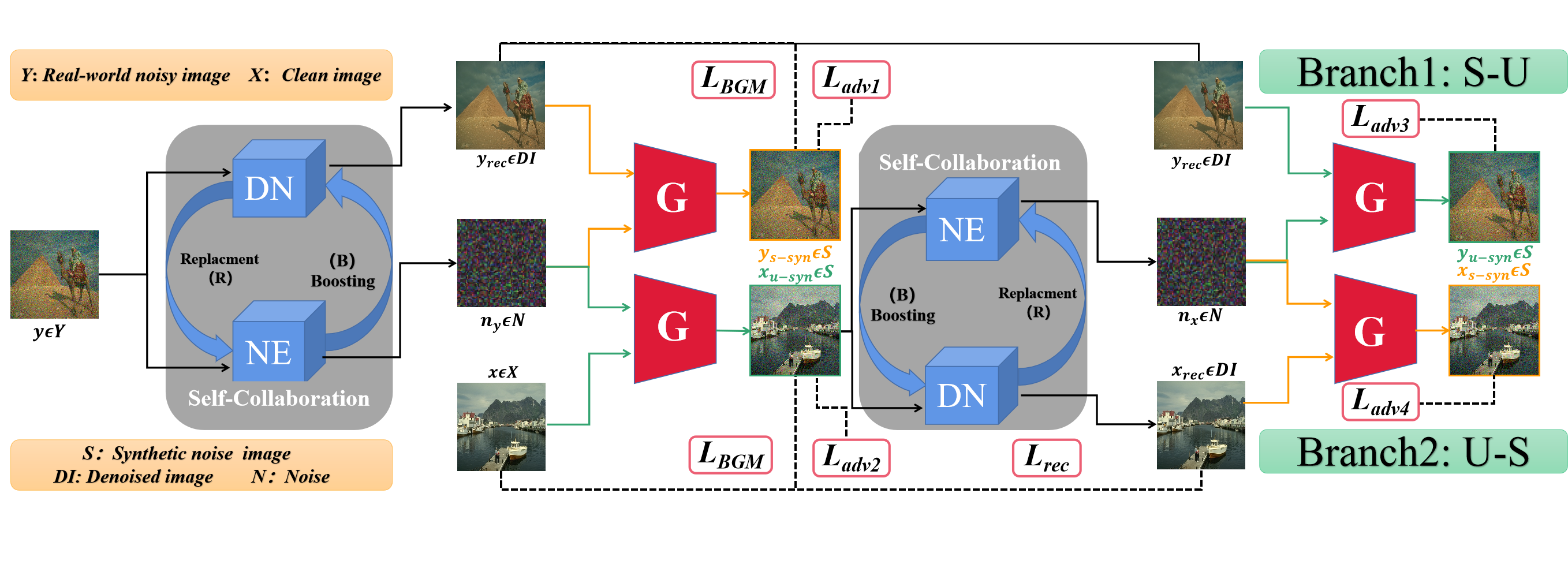}
\caption{\label{Fig2}
The architecture of our proposed SCPGabNet framework consists of two branches: Branch1: ``$\textcolor{red}{S}$elf-synthesis - $\textcolor{blue}{U}$npaired-synthesis" (Left: obtain \textcolor{red}{S}elf-synthesis image $y_{s-syn}$ from $y_{rec}$ and $n_{y}$; Right: obtain \textcolor{blue}{U}npaired-synthesis image $y_{u-syn}$ from $y_{rec}$ and $n_{x}$) and  Branch2: ``$\textcolor{blue}{U}$npaired-synthesis - $\textcolor{red}{S}$elf-synthesis" (Left: obtain \textcolor{blue}{U}npaired-synthesis image $x_{u-syn}$ from $x$ and $n_{y}$; Right: obtain \textcolor{red}{S}elf-synthesis image $x_{s-syn}$ from $n_{x}$ and $x_{rec}$). Each branch contains a self-collaborative operation that involves a noise extraction module \textbf{NE} and a denoiser \textbf{DN}. This process is essentially an \textbf{R} (replacement)-\textbf{B} (boosting) iteration, where the NE module extracts noise from the noisy image, and the DN removes the noise from the noisy image to generate a clean image.}
\end{figure*}
\section{Related Work}
\subsection{Denoising for Synthetic Noisy Image}
\par Image denoising techniques for synthetic noisy images, such as AWGN, can be classified into classical non-deep-learning-based, and deep-learning-based denoising approaches. Classical non-deep-learning-based approaches include filtering-based \cite{2, 3, 4, 5, 66}, model-based \cite{ 9, 32, 30, 12, 69, 61}, and traditional learning-based methods. However, the filtering-based methods require manually designed filters, and model-based methods require several preset hyperparameters and entail a significant computational burden. These limitations restrict their flexibility and efficacy. To address these challenges, DnCNN \cite{17} was proposed and achieved significant performance improvements over traditional methods. Subsequently, deep network-based methods \cite{17, 52, 53} became the mainstream in AWGN image denoising. For instance, several learning-based methods with advanced network designs have been proposed after DnCNN, including FFDNet \cite{54}, $N^{3}$Net \cite{21}, and MemNet \cite{53}. However, as pointed out by Guo et al. \cite{25}, AWGN-based training methods suffer in real-world denoising due to the domain differences between real and synthetic noise.
\subsection{Denoising for Real-World Image}
\par \noindent {\bf Real paired-dataset based supervised denoising methods.} These approaches typically involve designing high-performance denoising networks that are trained in a supervised manner using pairs of clean-noisy image datasets captured from real scenes.
Anwar et al. \cite{38} proposed a method that combines synthetic and real images during training to enhance the generality of the denoising model. Cheng et al. \cite{99} generated a set of image basis vectors from the noisy input images and reconstructed them from the subspace formed by these basis vectors to obtain image-denoising results. Ren et al. \cite{aind1} proposed a novel model-based denoising method that informs the design of the network for both synthetic and real denoising. Rencently, Ren et al. \cite{74} proposed a novel depth-unfolding network based on a latent space blind model via self-correction alternative optimization. 
\par \noindent{\bf Synthetic paired-dataset based two-step denoising methods.} While several real-world noisy datasets have been introduced, the limited number of training datasets motivates some approaches to divide the denoising task into two steps based on noise modeling: 1) synthesizing paired datasets by modeling noise or incorporating a priori noise information, and 2) training the denoiser. By leveraging prior knowledge of the image signal processing (ISP) pipeline, CBDNet \cite{25} emulated the pipeline using a gamma correction and demosaicking process. Then, the synthesized anisotropic Gaussian noise is transformed into a realistic noise signal that is used to generate training pairs for supervised learning. Liu et al. \cite{44} employed a priori information on image degradation to synthesize realistic noisy images and achieve good denoising performance. Zhou et al. \cite{pd} generated synthetic Gaussian-distributed noisy images and trained a Gaussian denoiser on a paired dataset of such images. For testing, they applied a conventional noise distribution conversion method that approximates real noise as Gaussian through pixel shuffling. Jang et al. \cite{jang} proposed a Clean-to-Noisy generator network based on a GAN that learns the features of real-world noises, capable of accurately representing various noise types. The generated pairs are then used to train a denoiser.
\par \noindent {\bf Image Denoising without Paired Dataset.} Obtaining paired datasets in practice is extremely challenging, which has led to the emergence of simple unsupervised methods. These methods do not require prior knowledge of the noise model but instead combine image synthesis and denoising within a single framework. Chen et al. \cite{chen} firstly proposed a noise generator that can create pseudo-noisy images to train a denoiser.  Cha et al. \cite{cha} proposed GAN2GAN, which uses a multi-generator/discriminator structure to better extract noisy information and generate pseudo images that better match the real noise distribution. Hong et al. \cite{hong} proposed UIDNet, which employs a sharpening processing mechanism to achieve noise separation and better train unpaired denoising models. In addition, several methods have been proposed to train models using only noisy images. Neshatavar et al. \cite{cvf} developed a self-supervised network that can handle real-world signal-dependent noise with greater adaptability to realistic noise. LEE et al. \cite{apbsn} proposed an unpaired learning approach that combines cyclic adversarial learning and self-supervised residual learning.
\section{Proposed Method}\label{sec3}
\par In this section, we introduce a promising denoising network called SCPGabNet. The details are illustrated in the following subsections.
\subsection{Parallel Generative Adversarial Branches for Unsupervised Denoising (PGabNet)}\label{sec3_1}
\par To apply the SC strategy, we first propose a high-performance baseline in this subsection, named PGabNet, which ensures the stability and effectiveness of the training process.
\subsubsection{Filter-Guided Synthetic Noisy Image Generator}
\label{3.1.1}
\par While supervised denoising frameworks generally outperform unsupervised ones (e.g., \cite{nafnet} 40.30dB, \cite{aind1} 39.35dB vs. \cite{cvf} 34.71dB, \cite{apbsn} 34.90dB on the SIDD Benchmark), they require a large number of paired datasets. Conversely, unsupervised denoising frameworks are more robust to situations where training images are inadequate or unpaired, compared to supervised methods. Acording to \cite{cha,jang}, the reason for this performancec gap is the difference between the synthetic noisy images and the real ones. The more realistic the training images are, the better the denoiser's performance will be. If the quality of the synthetic image is good enough (infinitely close to the real image), then the performance of the denoiser will not differ between supervised and unsupervised frameworks. Therefore, improving the quality of the synthetic image is a general method to enhance the performance of unsupervised denoising framework.
\par As depicted in Figure \ref{Fig3}, synthesizing real noisy images with high quality can be challenging due to the influence of different image contents. To mitigate this difficulty, we propose a novel filter-guided synthetic noisy image generator for our GAN model, which better captures noise information. Since it is difficult to learn the noise distribution directly, instead of directly inputting a noisy image and a clean image into $G$, we use the NE module to obtain the noise. Specifically, a learnable convolutional block called $DN_0$ is used for denoising, and the noise is then obtained by subtracting the denoised image from the original noisy image. This approach reduces the learning difficulty of the generator and encourages it to focus more on synthesizing high-quality noisy images, thereby improving the overall denoising performance.
\subsubsection{Parallel Generative Adversarial Branches}
\label{3.1.2}
\par The overall framework is a GAN structure that utilizes unpaired clean and noisy images. We employ a Resnet with 6 residual blocks as the generator, a PatchGAN as the discriminator \cite{patch} and a DeamNet \cite{aind1} as the denoiser. The specific structures of these variants are shown in detail in the \textcolor{blue}{Supplementary Material}.
\par There are two cases in learning synthetic pseudo noisy images   in PGabNet: (1) when the clean and noisy images are different; (2) when the clean and noisy images are the same. Unpaired synthesis is a common approach in many computer vision works, where one degraded image guides the generation of a synthetic image from another clean image (e.g., \cite{1000, jang}). The goal is to learn the noise distribution properties from the input noise of one noisy image to guide the generation of a similar pseudo-signal-dependent noisy image from another clean image. This method imposes unpaired constraints on the generator, which captures more prior information and improves the quality of synthetic noisy images. A robust generator should learn the real noise distribution properties of different inputs. To balance the noise extracted from ``same image" and ``different images", we propose self-synthesis. These two complementary constraints can improve the adversarial performance of the generator-discriminator and produce synthetic noisy images that are more consistent with the true noise distribution. 
\par As illustrated in Figure \ref{Fig2}, the PGabNet comprises two branches, with each branch implementing the complementary ``self-synthesis" and ``unpaired synthesis" constraints, respectively. Branch 1 utilizes ``self-synthesis - unpaired synthesis" architecture, while branch 2 employs the ``unpaired synthesis - self-synthesis" architecture. Specifically, branch 1 generates the self-synthesis noisy image $y_{s-syn}$ and the unpaired synthesis noisy image $y_{u-syn}$, while branch 2 generates the unpaired synthesis noisy image $x_{u-syn}$ and the self-synthesis noisy image $x_{s-syn}$. These images are then fed as inputs to the discriminator, along with the real noisy image $y$. The ``self-synthesis" constraint and the ``unpaired synthesis" constraint are strongly complementary within each branch of PGabNet. Additionally, the ``self-synthesis" constraint and the ``unpaired synthesis" constraint between these two branches are also complementary.
\begin{figure}[t]
\centering
\includegraphics[width=1\linewidth]{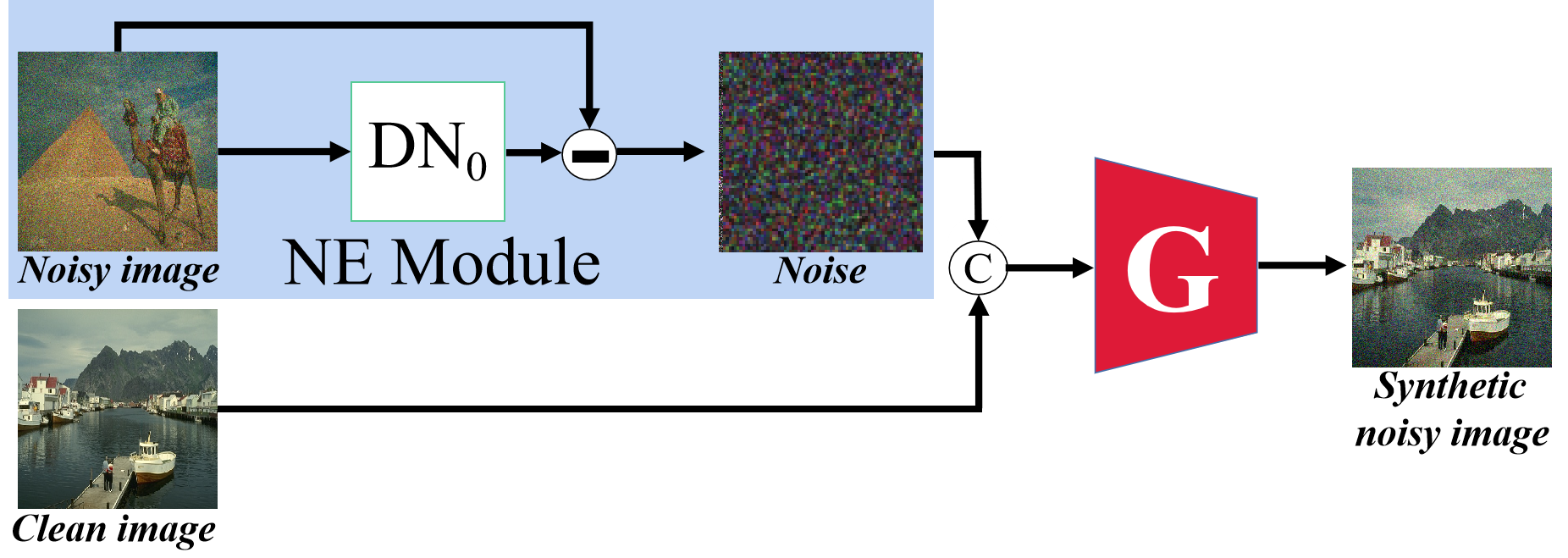}
\caption{\label{Fig3}
The architecture of the filter-guided synthetic noisy image generator is designed to extract real noise through the noise extraction (NE) module and project the noise distribution directly onto the clean image. This approach reduces the difficulty for the generator to synthesize better noisy images.}
\end{figure}
\subsubsection{GAN-based Noise Synthesize and Loss function}
\label{3.1.3}
As depicted in Figure \ref{Fig2}, $x$ and $y$ represent the clean and real-world noisy images. Correspondingly, $X$ and $Y$ denote the clean and real-world noisy domains. The generator $G$ aims to perform domain transformation by learning the image distribution in an unsupervised GAN framework. Simultaneously, the discriminator $D$ (Due to space limitation, it is not given in this figure. Please see \textcolor{blue}{Supplementary Material} for details.) evaluates whether the output of the generator belongs to the same domain as the target image. The generator and discriminator are trained in an adversarial manner to accomplish the domain transformation. Many studies \cite{jang, 25, ntgan} have demonstrated that signal-dependent noise can be modeled by additivity. The $G$ in Figure \ref{Fig3}, we first extract the noise $n_{y}$ from the noisy image using the NE module, and then input both the noise and clean image into $G$ to synthesize a pseudo noisy image:
\begin{equation}
x_{u-syn}=G(x,NE(y))
\end{equation}
\par We simultaneously train a discriminator $D$ to distinguish whether a given noisy image is synthesized by our generator $G$ or sampled from a real-world dataset. Here, $y$ and $x_{u-syn}$ denote the real-world image and synthetic image, respectively. To prevent model degradation during training and improve the representation capability of the network, we use the least squares loss as the adversarial loss for $L_{adv2}$. The mathematical expression is as follows:
\begin{equation}
\begin{split}
L_{adv2}(D,G)=&-E_{y \sim Y}[\Vert D(y)-1\Vert_{2}^{2}] \\
&-E_{x_{u-syn}\sim S}[\Vert D(x_{u-syn})-0\Vert_{2}^{2}]
\end{split} 
\end{equation}
\par That means for the generated image $x_{u-syn}$, its adversial loss $L_{adv2}$ is constrained between $y$ and $x_{u-syn}$. The other three dversial losses can be constructed similarly by constraining the current generated image and $y$. Then, the total loss can be given by:
\begin{equation}
L_{GAN} = L_{adv1}+L_{adv2}+L_{adv3}+L_{adv4}
\end{equation}
\par
\begin{figure}[t]
\centering
\includegraphics[width=1\linewidth]{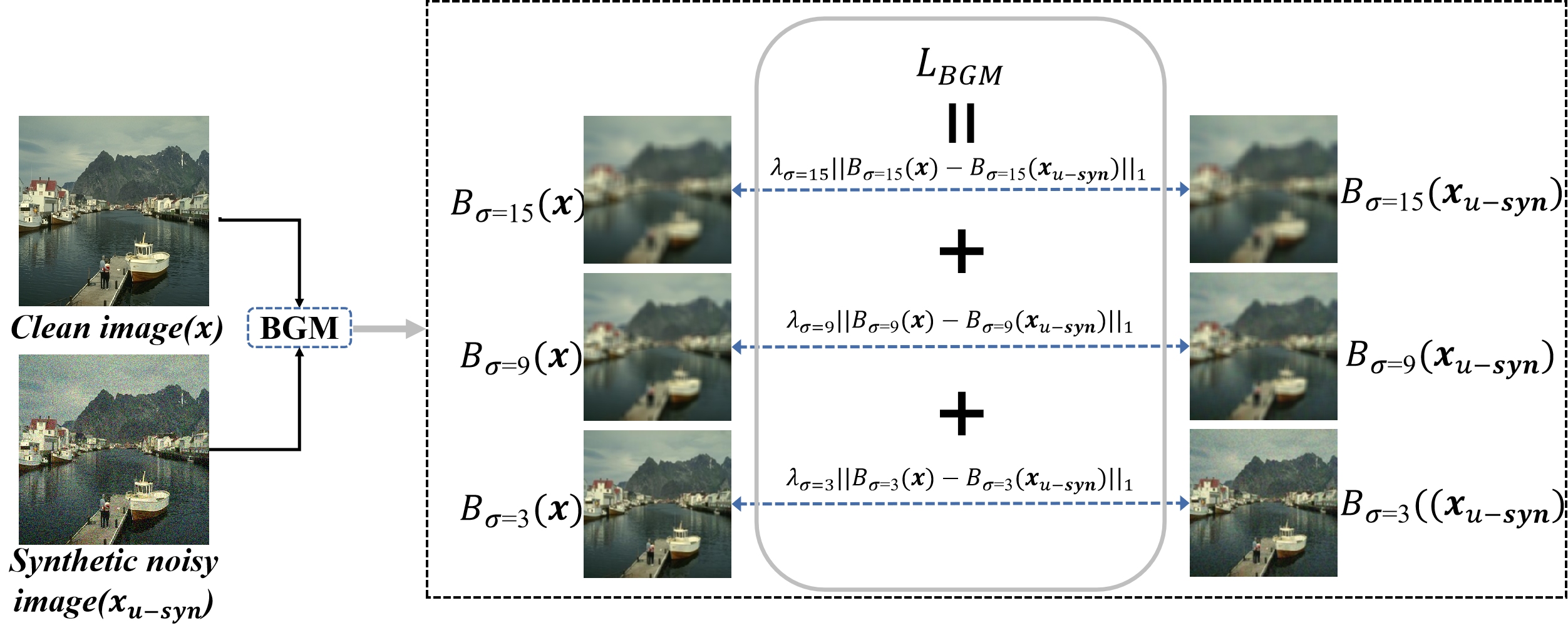}
\caption{\label{Fig5}
Example of BGM loss.}
\end{figure}

 Inspired by \cite{bgm}, we further apply a Background Guidance Module (BGM) to provide additional reliable supervision. The BGM maintains the consistency of the background between the synthetic noisy image and the clean image, constraining their low-frequency contents to be similar. Specifically, we illustrate this approach using $L_{BGM}$ in branch 2. Low-frequency contents are extracted by using several low-pass filters and constrained to be similar to each other through the L1 norm loss :
\begin{equation}
\begin{split}
\label{bgm}
L_{BGM}=&E_{x_{u-syn}\sim S,x\sim X} \\
&[\sum_{\sigma=3,9,15}\lambda_{\sigma}\Vert B_{\sigma}(x)-B_{\sigma}(x_{u-syn})\Vert_{1}]
\end{split} 
\end{equation}
\noindent where $B_{\sigma}$($\bullet$) denotes the Gaussian filter operator with blurring kernel size $\sigma$, and $\lambda_{\sigma}$ denotes the weight for the level $\sigma$. An example of the BGM loss is shown in Figure \ref{Fig5}. We empirically set $\sigma$-s to 3, 9 and 15, and $\lambda$-s to 0.01, 0.1, and 1, respectively.
\par In our image denoising framework, we utilize pseudo-paired samples denoted by $x_{i}$ and $({x}_{rec})_{i}$. The denoiser is trained by optimizing the following loss functions:
\begin{equation}
\begin{split}
\label{ssim}
L_{DN}(\Theta)=&\dfrac{1}{2m}\sum_{i=1}^m[\Vert ({x}_{rec})_{i}-x_{i}\Vert_{1} \\
&+ \lambda_{SSIM}L_{SSIM}(({x}_{rec})_{i},x_{i})]
\end{split} 
\end{equation}
\noindent where $m$ denotes the total number of the sample pairs, $({x}_{rec})_{i}$ is the clean image estimated by the denoising network, $L_{SSIM}$($\bullet$) represents the structural information used by SSIM loss to constrain the image, and $\lambda_{SSIM}$ is the weight for $L_{SSIM}$.
In conclusion, the total loss function is:
\begin{equation}
L = \mathop{min}\limits_{G}\mathop{max}\limits_{D}L_{GAN}+\lambda_{BGM}L_{BGM} + L_{DN}(\Theta)
\end{equation}
\noindent where $\lambda_{BGM}$ denotes the hyperparameter of background consistency loss.
\subsection{Proposed SC based PGabNet (SCPGabNet)}\label{sec3_3}
In the previous subsection, we introduced our powerful baseline,  PGabNet. However, as PGabNet is still a conventional GAN-based unsupervised framework, it is highly challenging to further improve its performance without significantly altering the network architecture or increasing inference complexity. To address this problem, we propose SC-based PGabNet (SCPGabNet).

\begin{figure}[t]
\centering
\includegraphics[width=1\linewidth]{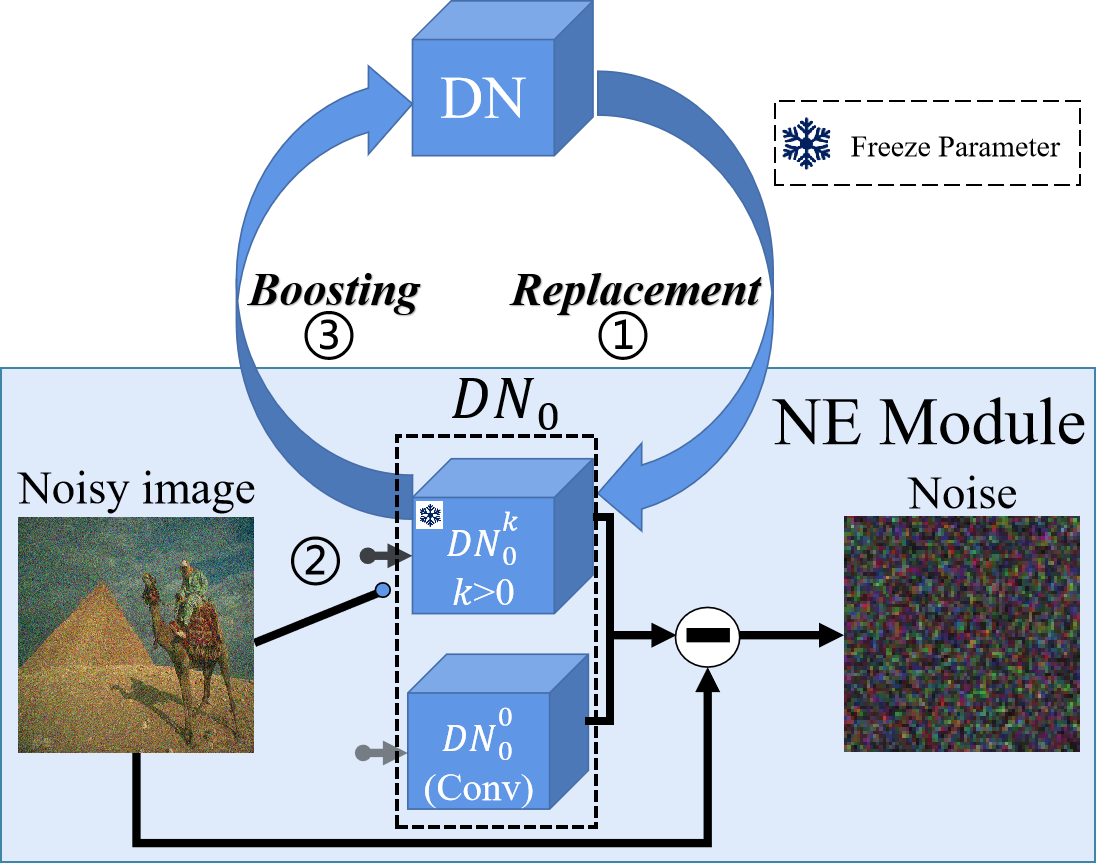}
\caption{\label{Fig4}
Visual illustration of the SC strategy. In \textbf{NE}, $k$ denotes the number of iterations. It is a learnable convolutional block when $k$=0. However, when $k$\textgreater0, the iterative collaboration of \textbf{$DN_{0}$} and the denoiser DN is conducted. Specifically, the weaker denoiser \textbf{$DN_{0}$} is successively replaced by the current more powerful denoiser DN, which enables the synthesis of better clean-noisy image pairs. This iterative process results in a series of increasingly powerful denoisers.
\textcircled{\scriptsize{1}}: the current denoiser DN is applied to replace the previous weaker denoiser $DN_{0}$. This replacement allows for the creation of a new and more powerful denoiser DN.
\textcircled{\scriptsize{2}}: the updated NE is obtained to generate better clean-noisy image pairs that are used to train the denoiser DN.
\textcircled{\scriptsize{3}}: the denoiser DN is trained using the updated clean-noisy image pairs, which further boosts its performance.}
\end{figure}

\subsubsection{Self-Collaboration Strategy}
\indent Our SC strategy is an innovative approach that enables a denoiser trained in conventional structures to self-correct and improve its performance without requiring modifications to the denoiser's structure or increased complexity. By incorporating the SC strategy into PGabNet, we aim to surpass the performance limit of conventional GAN-based unsupervised frameworks.

\par The SC strategy is illustrated in Figure \ref{Fig4}, and it functions more like an outer loop than an inner loop. The structure consists of a noise extraction module (NE) and a denoiser (DN). As described in the previous sub-sections, the NE module extracts noise from a noisy image and guides the generator to produce high-quality synthetic noisy images. To train the PGabNet, we initially use a simple and learnable linear convolutional layer as $DN_{0}^{0}$ in the NE module. 
Then, the denoiser is iteratively replaced and boosted, with $k$ denoting the number of iterations. During each iteration, the current more powerful denoiser DN replaces the previous weaker denoiser $DN_{0}$ in the NE module, significantly improving the performance of the updated denoiser. By using the SC strategy for the denoiser inside and outside the NE module, better $DN_{0}$ can be obtained, which can extract more accurate noise. This results in the production of more realistic synthetic noisy pairs, and iteratively improves the performance of the updated DN with higher-quality synthetic samples.  Excitingly, we observe a significant improvement in the performance of DN by using the SC strategy than the original one without SC.

During the SC stage, the loss functions of G and D are the same as before. The loss function of DN:

\begin{equation}
\begin{split}
\label{SC}
L_{DN-SC}= &L_{DN} + \dfrac{1}{2m}[\sum_{i=1}^m\Vert ({x}_{rec})_{i}-DN_{0}(x_{u-syn})\Vert_{1} \\
&+ \Vert ({y}_{rec})_{i}-DN_{0}(y_{i})\Vert_{1}) \\
&+\lambda_{SSIM}L_{SSIM}(({x}_{rec})_{i},DN_{0}(x_{u-syn})) \\
&+\lambda_{SSIM}L_{SSIM}(({y}_{rec})_{i},DN_{0}(y_{i}))] 
\end{split} 
\end{equation}

\par The basic idea of the SC is to use the result of the previous stage as feedback information to guide and improve the subsequent stages. The `feedback information' can either be a structure or an output image.  This approach has significant potential in low-level vision tasks.
\subsubsection{Analysis of SC for PGabNet} 
\indent To develop a more powerful denoising framework, we integrate the SC strategy with our previous PGabNet, which we now refer to as SCPGabNet. At the start of each iteration, the new NE can more accurately capture noise in the noisy image by replacing  $DN_{0}$ in the NE module with a more powerful DN. This reduces the impact of the image content on the synthetic noisy image generation process. As illustrated in Figure \ref{Fig2}, with more precise noise extracted from the noisy image $y$, our denoiser can achieve better results in both self-synthesis in branch 1 and unpaired synthesis in branch 2, resulting in a higher 
quality synthetic noisy image. Similarly, more accurate noise extracted from the synthetic noisy image $x_{u-syn}$  can improve unpaired synthesis in branch 1 and self-synthesis in branch 2, thus promoting complementary constraints between the two branches and making the modules in the network more interconnected. As a result, our SC strategy creates a self-boosting framework that enables better denoiser training and performance.
\par The implementation of SC strategy in SCPGabNet involves several steps: First, after the original framework (PGabNet) has reached  convergence, we replace $DN_{0}$ in the NE module with the latest DN and fix its parameters to generate better pseudo-noisy images. Next, we retrain G, D, and DN until convergence is achieved. Finally, we repeat the previous processes until the performance of DN no longer improves.
\begin{table*}
\begin{center} 
\fontsize{7}{7}\selectfont
\caption{\label{tab1} Denoising results of several competitive methods on SIDD Validation, SIDD Benchmark, and DND Benchmark} 
\setlength{\tabcolsep}{3.3mm}{
\begin{tabular}{ccccccccc}
\toprule
&\multirow{2}[3]{*}{Methods} & \multirow{2}[3]{*}{Conference/Journal} &\multicolumn{2}{c}{SIDD Validation} 
& \multicolumn{2}{c}{SIDD Benchmark} & \multicolumn{2}{c}{DND Benchmark } \\
\cmidrule{4-9} & & & PSNR$\uparrow$ & SSIM$\uparrow$ & PSNR$\uparrow$ & SSIM$\uparrow$ & PSNR$\uparrow$ & SSIM$\uparrow$\\
\midrule
\multirow{2}{*}{Non-learning} & BM3D \cite{5} & TIP 2007 & 31.75 & 0.7061 & 25.65 & 0.6850 & 34.51 & 0.8510 \\
& WNNM \cite{12}& CVPR 2014 & $-$ & $-$ & 25.78 & 0.8090 & 34.67 & 0.8650 \\
\midrule
\multirow{9}{*}{Real pairs( Supervised)}& TNRD \cite{51} & TPAMI 2016 & 26.99 & 0.7440& 24.73 & 0.6430 & 33.65 & 0.8310 \\
& DnCNN \cite{17} & TIP2017 & 26.20 & 0.4414 & 28.46 & 0.7840 & 32.43 & 0.7900 \\
& FFDNet \cite{54} &TIP 2018 & 26.21 & 0.6052 & 29.30 & 0.6940 & 34.40 & 0.8470 \\
& RIDNet \cite{ridnet}& CVPR 2019 & 38.76 & 0.9132 & 37.87 & 0.9430 & 39.25 & 0.9530 \\
& AINDNet \cite{aind2} & CVPR 2020 & 38.96 & 0.9123 & 38.84 & 0.9510 & 39.34 & 0.9520 \\
& InvDN \cite{44} & CVPR 2021 & 38.30 & 0.9064 & 39.28 & 0.9550 & 39.57 & 0.9520 \\
& DeamNet \cite{aind1} & CVPR 2021 & 39.40 & 0.9169 & 39.35 & 0.9550 & 39.63 & 0.9531 \\
& ScaoedNet \cite{74} & NeurIPS 2022 & 39.52 & 0.9187 & 39.48 & 0.9570 & 40.17 & 0.9597 \\
\midrule
\multirow{5}{*}{Synthetic pairs(two pipeline)}& DnCNN \cite{17} & TIP 2017 & $-$ & $-$ & 23.66 & 0.5830 & 32.43 & 0.7900 \\
& CBDNet \cite{25} & CVPR 2019 & 30.83 & 0.7541 & 33.28 & 0.8680 & 38.06 & 0.9420 \\
& PD+ \cite{pd} & AAAI 2020 & 34.03 & 0.8810 & 34.00 & 0.8980 & 38.40 & 0.9450 \\
& C2N$+$DnCNN \cite{jang}& ICCV 2021& $-$ & $-$ & 33.76 & 0.9010 & 36.08 & 0.9030 \\
& C2N$+$DIDN \cite{jang}& ICCV 2021 & $-$ & $-$ & 35.02 & 0.9320 & 36.12 & 0.8820 \\
\midrule
\multirow{9}{*}{Unsupervised}& N2V \cite{n2v} & CVPR 2019 & 29.35 & 0.6510 & 27.68 & 0.6680 & $-$ & $-$ \\
& GCBD \cite{chen} & CVPR 2018 & $-$ & $-$ & $-$ & $-$ & 35.58 & 0.9220 \\
& UIDNet \cite {hong} & AAAI 2020 & $-$ & $-$ & 32.48 & 0.8970 & $-$ & $-$ \\
& R2R \cite{r2r} & CVPR 2021 & 35.04 & 0.8440 & 34.78 & 0.8980 & 36.20 & 0.9250 \\
& CVF-SID ($S^{2}$) \cite{cvf}& CVPR 2022 & $-$ & $-$ & 34.71 & 0.9170 & 36.50 & 0.9240 \\
& AP-BSN\cite{apbsn} & CVPR 2022 & 34.46 & 0.8501 &34.90 & 0.9000 & 37.46 & 0.9240 \\
& \textbf{PGabNet(baseline)} & \textbf{$-$} & \textbf{34.66} & \textbf{0.8517} & \textbf{34.67} & \textbf{0.8950} & \textbf{36.87} & \textbf{0.9267} \\
& \textbf{SCPGabNet(ours)} & \textbf{$-$} & \textbf{36.53} & \textbf{0.8860} & \textbf{36.53} & \textbf{0.9250} & \textbf{38.11} & \textbf{0.9393} \\
\bottomrule
\label{table0}
\end{tabular}}%
\end{center}
\end{table*}%

\section{Experiments}
\par 
In this section, we first describe the experimental settings, including the datasets and training details. After that, to evaluate the effectiveness, the proposed method is compared with some representative supervised/unsupervised methods. Finally, we analyze the proposed method in-depth. 
\subsection{Experimental Setting}\label{sec4_1}
\par \textbf{Training and Testing Data.} To train and test the model,  

\noindent we first equally divide the SIDD Medium training set (consisting of 320 pairs of noise images and corresponding clean images captured by multiple smartphones) into separate noisy and clean image parts. Then, we use 160 clean images from the first part and 160 noisy images from the second part to construct an unpaired dataset of real images for training the algorithm presented in this paper. We evaluate the denoising performance on three widely used real-world noisy datasets SIDD\cite{sidd} Validation, SIDD\cite{sidd} Benchmark, and the DND\cite{dnd} Benchmark. Note that the denoised images of SIDD Benchmark and DND Benchmark can be uploaded to the SIDD and DND websites to obtain the PSNR and SSIM results.
\par \textbf{Implementation Details.} To optimize the proposed network, we adopt the Adam optimizer algorithm with $\beta_1$=0.9, $\beta_2$=0.999, and the initial learning rate is set to $10^{-4}$. The mini-batch size is set to 10, the used framework is PyTorch, and the used GPU is GeForce RTX 3090. For noise learning, the background consistency loss hyperparameter $\lambda_{BGM}$ in the loss function of Eq. \ref{bgm} is set to 6. For the denoising network, the hyperparameter $\lambda_{SSIM}$ of the SSIM constraint term in Eq. \ref{ssim} is set to 1.
\subsection{Real-World Image Denoising Analysis}\label{sec4_2}
\par In this subsection, we evaluate the denoising performance of our method on real images from the SIDD Validation, SIDD Benchmark, and DND Benchmark. We compare our method with a range of traditional-based methods, representative supervised methods based on paired images, and latest unsupervised methods based on unpaired images from recent years. We utilize evaluation metrics (PSNR and SSIM) to assess the effectiveness of each method.
\begin{figure*}[t]
\centering
\includegraphics[width=1\linewidth]{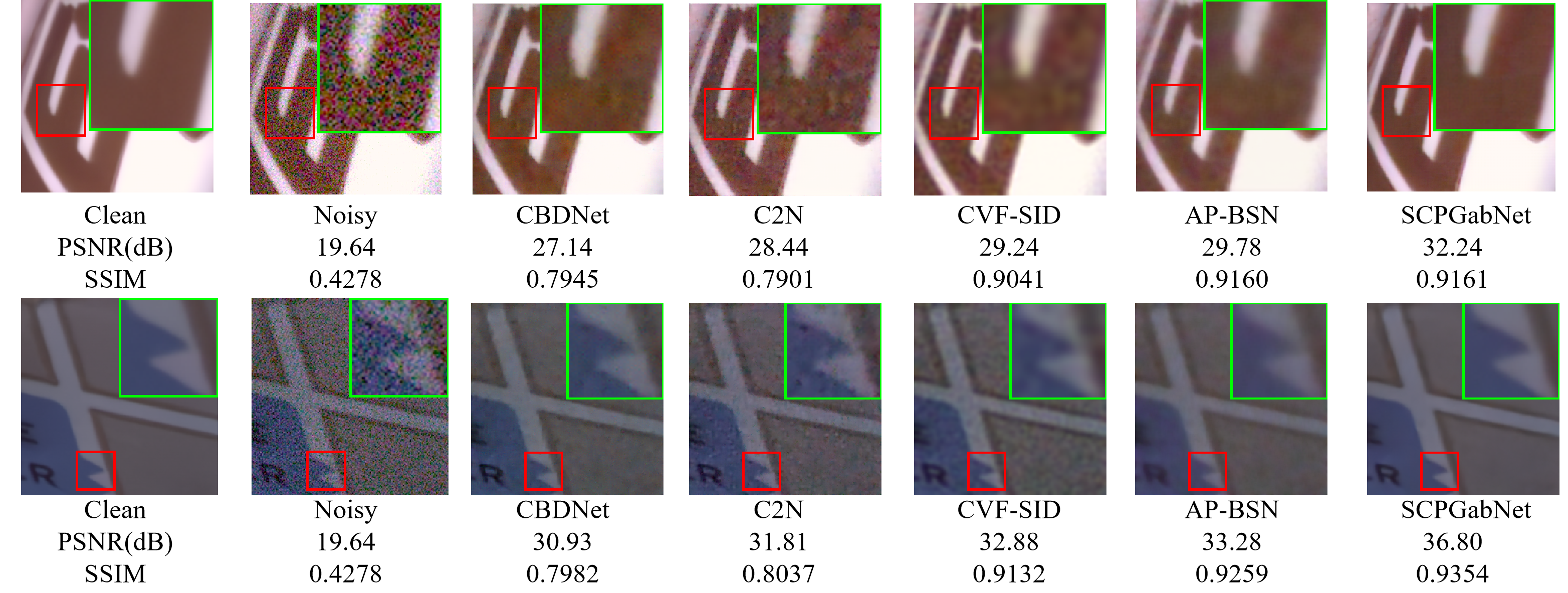}
\caption{\label{Fig8}
Visual comparison of our method against other competing methods. The quantitative PSNR(dB)/SSIM results are listed as well.}
\end{figure*}
\par Table \ref{table0} presents the quantitative evaluation of various methods on the SIDD Validation, SIDD Benchmark, and DND Benchmark in the sRGB space. It is observed that SCPGabNet achieves the best performance among the unsupervised approaches.
Specifically, compared with the latest unpaired methods AP-BSN and CVF-SID presented at CVPR 2022, SCPGabNet provides a PSNR gain of 1.63 dB and 1.82 dB, and an SSIM gain of 0.25 and 0.008 on the SIDD benchmark. Moreover, SCPGabNet outperforms CBDNet, which is a classic supervised real denoising network. In terms of the two-stage synthetic denoising method, SCPGabNet outperforms two-stage pipeline methods with self-ensemble (e.g., C2N+DnCNN$^*$, C2N+DIDN$^*$) on both SIDD and DND datasets. Although PD+\cite{pd} performs slightly better than SCPGabNet on the DND Benchmark, SCPGabNet's performance is significantly superior to PD+ on both SIDD Benchmark and SIDD validation by more than 2dB. SCPGabNet is trained only on the SIDD training set and still exhibits a greater advantage on other benchmarks, indicating its greater generalizability than two-step methods. Although our method's denoising performance is not as good as some of the latest supervised methods based on real image pairs, such as ScaoedNet\cite{74}, the majority of these methods depend on a large number of paired images. Consequently, they may not be effectively applied to real-world image-denoising tasks when training images are insufficient or cannot be paired. In contrast, our SCPGabNet can achieve image denoising without using paired images, making it flexible enough to handle a wide range of real-world denoising scenarios. Figure \ref{Fig8} depicts denoised  examples on the SIDD Validation dataset. Our method outperforms comparable unsupervised and two pipeline methods in noise removal and detail preservation.
\subsection{Ablation Study}\label{sec4_3}
\par In this subsection, we demonstrate the importance of the PGabNet structure and BGMloss in Table \ref{table1}. V1 is the GAN-based unsupervised denoising network with only unpaired synthesis. V2 is the V1 + BGMloss, V3 is the V2 + NE module. V4 is the branch U-S, i.e., SGabNet and V5 is our baseline (PGabNet). Due to the limited space, the specific structure of these variants are provided in the \textcolor{blue}{Supplementary Material}.
\begin{table}
\begin{center} 
\fontsize{8}{8}\selectfont
\caption{\label{tab1} $\textbf{V1}$: (U) Conventional GAN-based unsupervised denoising network only with unpaired synthesis; $\textbf{V2}$: V1 + BGMloss; $\textbf{V3}$: V1 + BGMloss + NE module; $\textbf{V4}$: (S) SGabNet(V1 + BGMloss + NE module + self-synthesis) $\textbf{V5}$: (P) PGabNet(our baseline).} 
\setlength{\tabcolsep}{3.3mm}{
\begin{tabular}{ccccccccc}
\toprule
Methods & V1 & V2 & V3 & V4 & \textbf{V5(ours)} \\
\midrule
U & \checkmark & \checkmark & \checkmark & \checkmark & \checkmark \\
BGMloss & & \checkmark &\checkmark & \checkmark & \checkmark \\
NE module & & & \checkmark & \checkmark & \checkmark \\
S & & & & \checkmark & \checkmark \\
P & & & & & \checkmark \\
\midrule
PSNR(dB) & 33.14& 33.26& 33.45& 34.27& 34.67 \\
\midrule
\end{tabular}}%
\label{table1}
\end{center} 
\end{table}%

\par By comparing V1 with V2, we found that adding the BGMloss to the GAN-based unsupervised denoising network can lead to a more stable training process. Furthermore, adding the NE module to V2 to obtain V3, the PSNR increase is about 0.20dB, verifying the effectiveneess of the NE module and the improvement in the quality of synthetic images. Comparing V3 with V4 (SGabNet), we observed that the denoiser experienced a more significant improvement after adding the ``self-synthesis" constraint. Specifically, the PSNR improved by 0.82dB on the SIDD Benchmark, demonstrating that ``self-synthesis + unpaired synthesis" can better train the network and enhance the denoiser's performance. Finally, compared PGabNet with SGabNet, we observed a significant performance improvement in PGabNet with PSNR gains of 0.40dB on the SIDD Benchmark. This is because PGabNet enhances PGabNet with SGabNet, leading to the production of better synthetic noisy images that are more consistent with the real image distribution and improving the denoiser's performance.\vspace{-0.3em}
\subsection{Effectiveness of SC strategy}\label{sec4_4}
\par
\begin{figure}[t]
\centering
\includegraphics[width=1\linewidth]{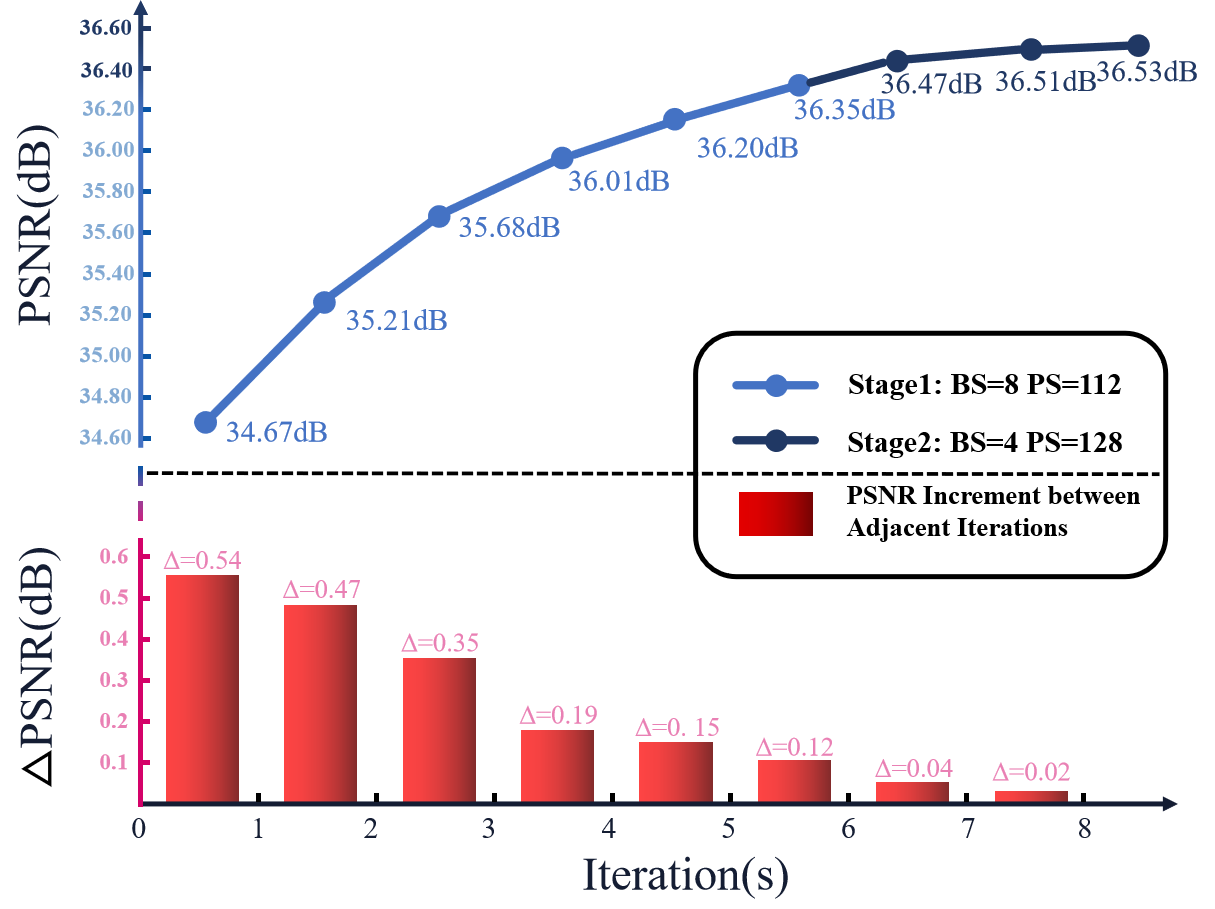}
\caption{\label{Fig11}
The entire process involves eight iterations, consisting of two stages. In the first stage: the batchsize=8 and the patchsize=112. In the second stage,  fine-tuning is performed with a batch size of 4 and a patch size of 128. The upper part of the figure shows the results of SCPGabNet on the SIDD Benchmark after each iteration, while the lower part displays the improvement achieved in each iteration.}
\end{figure}

 In this subsection, we apply the SC strategy to our baseline method (PGabNet) and obtain SCPGabNet. The details of this process are as follows: In the first stage, the batch size is 8 and the patchsize is 112. In the second stage, we increase the patchsize to 128 and reduce the batchsize to 4 for fine-tuning. The learning rate is keep at $10^{-4}$ for both stages. As shown in Figure \ref{Fig11}, we display the performance of the SC on the SIDD Benchmark. In the initial few iterations, we observe a notable improvement in denoising performance, especially in the first iteration where the denoiser improves by more than 0.5dB. Based on the PSNR increments observed between adjacent iterations, the $\Delta$PSNR decreases as the iterations progress, with the last iteration resulting in only a 0.02dB improvement. By comparison with PGabNet, with a PSNR of 36.53dB, significantly improved by 1.86dB on the SIDD Benchmark. This improvement highlights the effectiveness of our proposed approach in achieving state-of-the-art performance in image denoising.
\subsection{Analysis on Transferability}\label{sec4_5}
To evaluate the generality and transferability of our approach, we apply the proposed SC strategy to several classical and the latest denoising networks, including DnCNN \cite{17}, UNet \cite{unet}, DeamNet \cite{aind1} and DBSNL \cite{dbsnl}, which serves as the denoisier for the AP-BSN \cite{apbsn}. The framework still adopts the PGabNet proposed in section \ref{sec3} with different denoiser networks.
\par Table \ref{table4} illustrates that our proposed SC strategy is effective to these networks. For instance, after applying the SC strategy to DnCNN, the PSNR/SSIM improvements on SIDD Validation are 0.96dB/0.0085. For UNet and DBSNL, the gains are 1.58dB/0.0373 and 1.52dB/0.0035 respectively. Consequently, the results demonstrate that our SC strategy has strong transferability and can be potentially applied to other unsupervised denoising methods.
\begin{table}[htbp]
\fontsize{6}{6}\selectfont
\caption{\label{tab4} The effects of SC for different denoisers.}
\setlength{\tabcolsep}{1mm}{
\begin{tabular}{cccccccc}
\toprule
\multirow{2}[3]{*}{Networks} & \multirow{2}[3]{*}{Self-Collaboration} & \multicolumn{2}{c}{SIDD Validation}& \multicolumn{2}{c}{SIDD Benchmark} & \multicolumn{2}{c}{DND Benchmark } \\
\cmidrule{3-8} & & PSNR$\uparrow$ & SSIM$\uparrow$ & PSNR$\uparrow$ & SSIM$\uparrow$ & PSNR$\uparrow$ & SSIM$\uparrow$ \\
\midrule
\multirow{2}{*}{DnCNN \cite{17}}& No & 30.60 & 0.8553 & 30.56 & 0.8150 & 29.79 & 0.7876\\
&Yes & 31.56 & 0.8638 & 31.52 & 0.8260 & 30.77 & 0.7892\\
\midrule
\multirow{2}{*}{UNet \cite{unet}}& No & 34.46 & 0.8417 & 34.63 & 0.8960 & 35.81 & 0.9140\\
&Yes & 36.04 & 0.8790 & 36.00 & 0.9180 & 38.04 & 0.9401\\
\midrule
\multirow{2}{*}{DBSNL \cite{dbsnl}}& No & 34.38 & 0.8757 & 34.29 & 0.9100 & 36.51 & 0.9257\\
&Yes & 35.90 & 0.8792 & 35.86 & 0.9160 & 37.72 & 0.9295\\
\midrule
\multirow{2}{*}{Deamnet \cite{aind1}}& No & 34.66 & 0.8517 & 34.67 & 0.8950 & 36.87 & 0.9267\\
&Yes & 36.53 & 0.8860 & 36.53 & 0.9250 & 38.11 & 0.9393\\ 
\bottomrule
\end{tabular}}%
\label{table4}
\end{table}%
\section{Conclusion}\label{sec5}
In this paper, we first introduce a parallel generative adversarial branches for unsupervised real-world image denoising as our baseline. Furthermore, we innovatively propose an SC strategy that can provide the denoiser a self-boosting capacity and significantly improve denoising performance. Our experimental results demonstrate that the proposed method achieves state-of-the-art performance. In addition, we validate the transferability of the SC strategy on various denoisers and anticipate its potential applicability to many low-level computer vision tasks.

\section{Limitations and Future Work}\label{sec6} 

Our current focus is on unsupervised tasks, with supervised tasks slated for future exploration. We specifically investigate the denoising task and, at the same time, consider the potential to delve into restoration tasks. In the SC strategy, each iteration necessitates manually selecting the optimal model iteration or retraining within the phase based on metrics, which is labour-intensive. The SC strategy can be simplified in the future by using automated iteration. The overall framework is based on CNNs, and better performance can be obtained in the future with transformers.

\textbf{Acknowledgement} This work was supported by the National Natural Science Foundation of China under Grant 62171304.

{\small
\bibliography{e}
}

\end{document}